\documentclass{PoS}

\usepackage{bbm,amsmath} 
\usepackage{wrapfig}

\newcommand{\be}{\begin{equation}}
\newcommand{\ee}{\end{equation}}
\newcommand{\bea}{\begin{eqnarray}}
\newcommand{\eea}{\end{eqnarray}}

\newcommand{\bi}{\begin{itemize}}
\newcommand{\ei}{\end{itemize}}

\title{Corrections to the Banks-Casher relation with Wilson quarks}

\ShortTitle{Banks-Casher relation}

\author{
\vspace*{-5mm}
\begin{flushright}
CERN-PH-TH/2013-016
\end{flushright}
Silvia Necco \\
        CERN, Physics Departement, 1211 Geneva 23, Switzerland}
\author{\speaker{Andrea Shindler}\thanks{Heisenberg fellow}\\
        CERN, Physics Departement, 1211 Geneva 23, Switzerland\\
        E-mail: \email{andrea.shindler@cern.ch}}

\abstract{The Banks-Casher relation links the spectral density of the Dirac operator with the
existence of a chiral condensate and spontaneous breaking of chiral symmetry.
This relation receives corrections from a finite value of the quark mass,
a finite space-time volume and, if evaluated on a discrete lattice, from the finite value of the
lattice spacing $a$. We present a status report of a determination of these corrections for Wilson quarks.}

\FullConference{The 7th International Workshop on Chiral Dynamics,\\
		August 6 -10, 2012\\
		Jefferson Lab, Newport News, Virginia, USA}

\begin{document}
\section{Introduction}
\vspace{-0.3cm}
\noindent The Banks-Casher relation~\cite{Banks:1979yr} relates the spectral density $\rho_D$
of the Hermitian Dirac operator $-iD$ with the chiral condensate $\Sigma$
\be
\rho_D\left(\gamma,m\right) = \frac{\Sigma}{\pi}\left[1+O(\{|\gamma|,m\}/\Lambda_{QCD}) \right]\,,
\label{eq:BC}
\ee
where $\gamma_k$ are the eigenvalues of the massless operator and $m$ is the sea quark mass.
In principle the Banks-Casher relation provides a tool to determine the 
chiral condensate~\cite{Giusti:2008vb} with a lattice QCD computation.

A more standard way to determine the chiral condensate with lattice QCD computations is to study
the quark mass dependence of the pion mass and comparing it with the predictions of 
chiral perturbation theory ($\chi$PT).
Recent lattice calculations are performed close to the physical values of the
quark masses (see ref.~\cite{Jung:2010jt,Wittig:2012np} for recent reviews).
To control accurately the light quark mass
dependence of hadronic quantities it is important, if possible, 
to have independent determinations of leading order (LO) low energy constants (LECs),
as for example the chiral condensate. 
\begin{wrapfigure}{r}{0.45\textwidth}
  \vspace{-20pt}
  \begin{center}
    \includegraphics[width=0.43\textwidth]{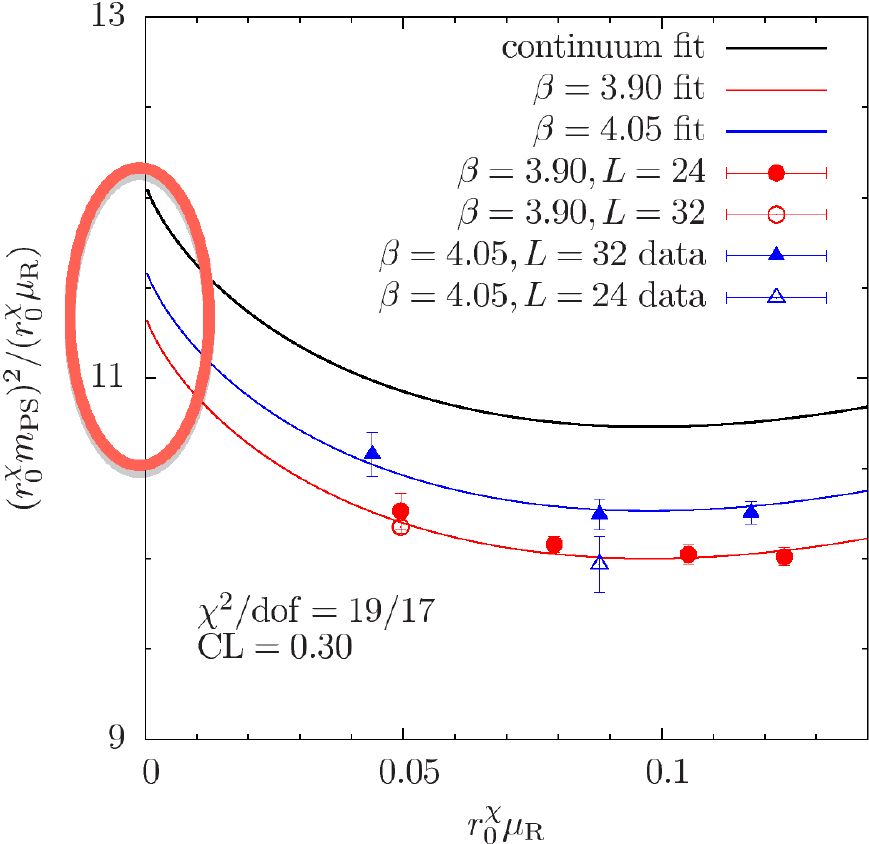}
  \end{center}
  \vspace{-20pt}
\caption{Light quark mass dependence of the ratio between the pseudoscalar meson mass squared and the renormalized 
light quark mass in units of the Sommer parameter $r_0$.}
  \vspace{-10pt}
\label{fig:motivations}
\end{wrapfigure}
In the plot in fig.~\ref{fig:motivations} we show the light quark mass dependence of the squared pion mass
computed by the European Twisted Mass Collaboration (ETMC) with $N_f=2$ dynamical light quarks~\cite{Baron:2009wt}.
Despite the rather good description of the lattice results with $\chi$PT, it is obviously desirable
being able to predict, in an independent way, the behaviour close to the chiral limit (red ellypse)
to better constrain the light quark mass dependence.
A constraint of the chiral fit will be beneficial for a more accurate determination
of the next-to-leading order LECs and a better confidence on the chiral fits.

An example of independent determination of LO LECs is provided by lattice 
calculations in the so called epsilon-regime, where first estimates of the 
chiral condensate and decay constant are rather encouraging~\cite{Jansen:2008ru,Hasenfratz:2008ce}.
Using Wilson-type fermions as a QCD discretization, calculations are affected by potentially large cutoff effects, 
thus it is important to have a theoretical analysis of quantities such as the spectral density,
based on Wilson chiral perturbation theory (W$\chi$PT)~\cite{Sharpe:1998xm,Bar:2003mh}.~\footnote{The analysis of two-point functions in the epsilon-regime within the framework of W$\chi$PT
has been discussed in refs.~\cite{Shindler:2009ri,Bar:2008th} 
and more recently extended in ref.~\cite{Akemann:2012bc}.}
These proceedings are a status report of an ongoing attempt to understand cutoff and finite size
effects affecting the spectral density of the Wilson operator. 
\vspace{-0.4cm}
\section{Chiral condensate from the mode number}
\label{sec:pregime}
\vspace{-0.4cm}
\noindent With Wilson fermions it is advantageous to consider the Hermitean Wilson-Dirac operator $Q=\gamma_5 D_m$,
where $D_m$ is the massive Wilson opearator.
To compute the spectral density $\rho_Q$ in W$\chi$PT
one introduces a flavour doublet of valence fermions $\chi_v$ with a Wilson twisted mass
action and twisted mass $\mu_v$.
\begin{figure}[tb]
\vspace{-0.9cm}\includegraphics[width=0.43\textwidth,angle=0]{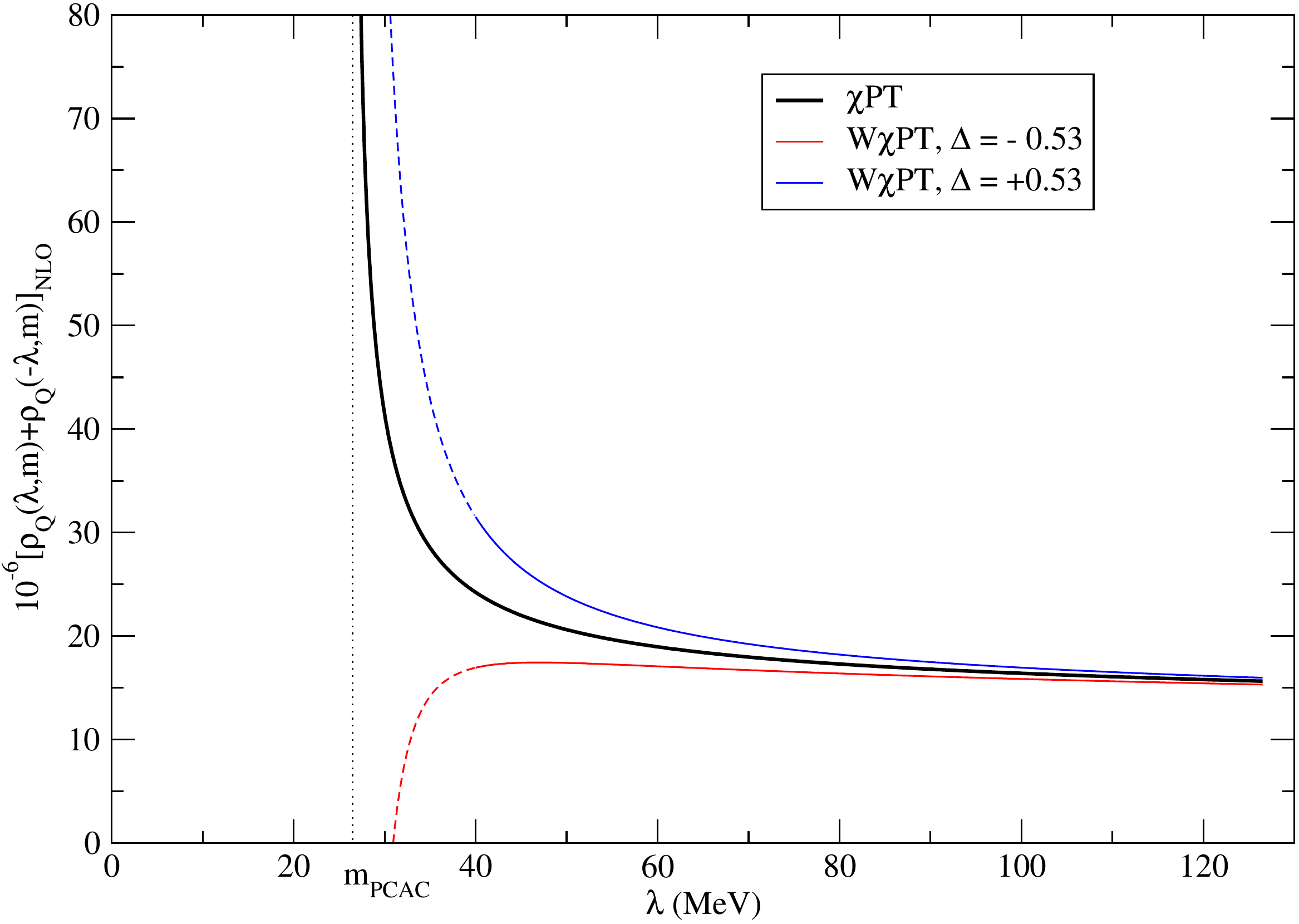}
\hbox{\raise -5mm\hbox{\includegraphics[width=0.51\textwidth,angle=0]{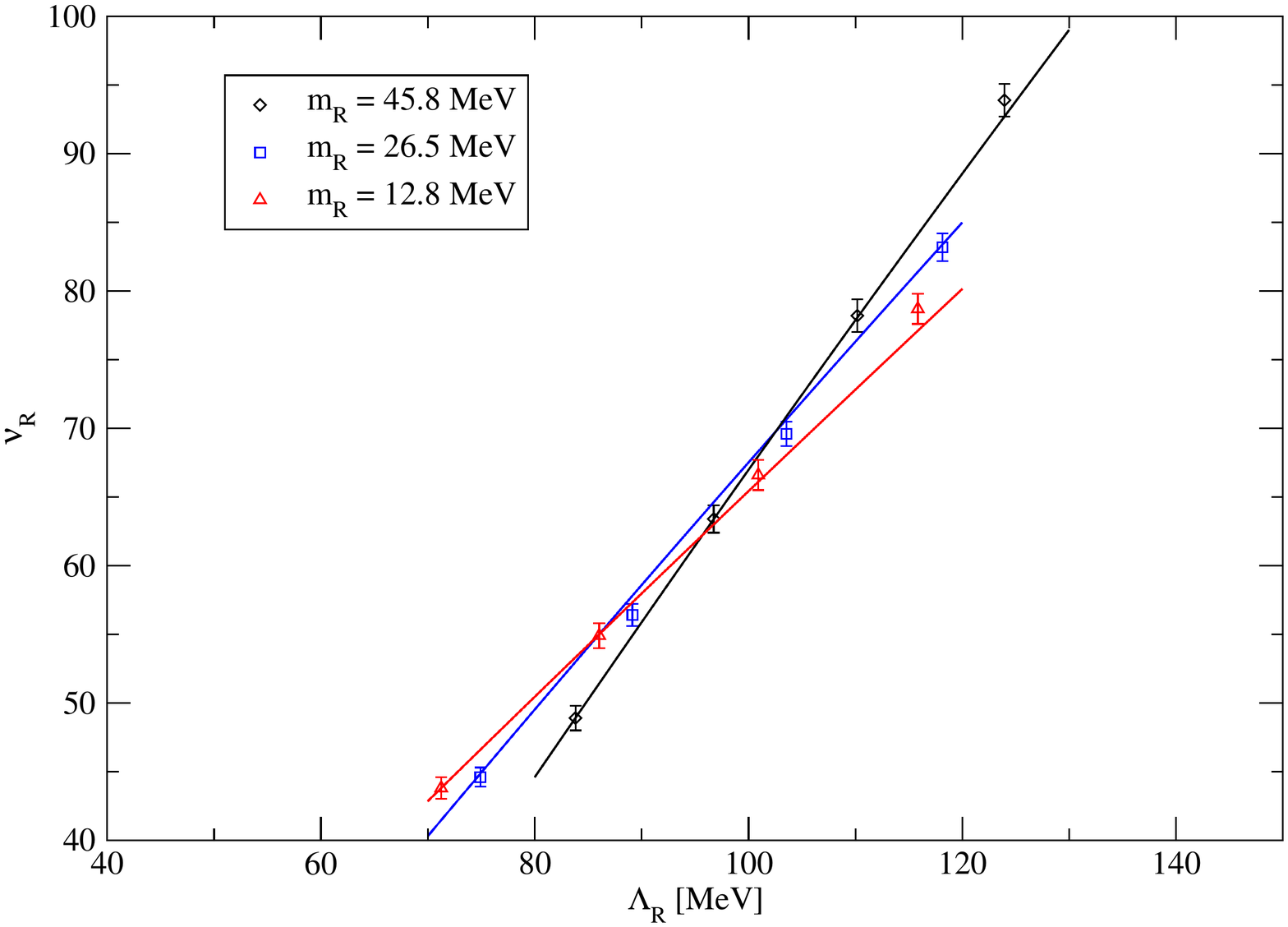}}}
\vspace{-0.5cm}
\caption{Left plot: the spectral density $[\rho_Q(\lambda,m)+\rho_Q(-\lambda,m)]_{NLO}$ in the infinite volume. 
We used the parameters $\Sigma=(275\;{\rm MeV})^3$, $m_{\rm PCAC}=26.5$ MeV, $F=90$ MeV, $\bar{L}_6=5$, $\mu=139.6$ MeV. 
The solid black line corresponds to the continuum $\chi$PT prediction, while 
the red (blue) lines correspond to the lattice W$\chi$PT prediction ($O(a)$-improved) 
on eq.~\protect\ref{eq:rhoQ_nlo} with $\hat{a}^2W'_8= \pm 5\cdot 10^6\; {\rm MeV}^4$, 
corresponding to $\Delta=\mp 0.53$. Right plot: result of the global fit of the data published in~\cite{Giusti:2008vb} with our formula for the 
renormalised mode number (eqs.~\protect\ref{eq:mn} and~\protect\ref{eq:rhoQ_nlo}). The fit parameters are $\Sigma$, 
$\Delta$ and $\overline{L}_6$ and we obtain
$\chi^2/{\rm dof} = 0.91$. 
}
\vspace{-0.3cm}
\label{fig:rhonu_p}
\end{figure}
The spectral density is related to the discontinuity of the valence pseudoscalar
condensate along the imaginary axis in the twised mass plane~\cite{Sharpe:2006ia}
\be
{\rm Disc}\left.\left[\langle \bar\chi_v\gamma_5\tau^3\chi_v\rangle \right]\right|_{\mu_v=i\lambda} = 
2i\pi\left[\rho_Q(\lambda,m) + \rho_Q(-\lambda,m)\right]\,,
\label{eq:disc_P3}
\ee
thus the valence pseudoscalar condensate is a tool to compute the spectral density in W$\chi$PT
\footnote{We recall that in the continuum the two spectral densities are connected by the relation 
$\rho_Q(\lambda,m) = \frac{\lambda}{\sqrt{\lambda^2-m^2}}\rho_D\left(\sqrt{\lambda^2-m^2},m \right)$.}.
After matching the continuum Symanzik effective theory with the generalized effective chiral Lagrangian
one needs to choose a proper power counting for the scales involved in the problem.
The scales are the sea quark mass, $m$, the valence twisted mass $\mu_v$
(directly related to $\lambda$), the lattice spacing $a$
and the linear size of the space-time volume $L$.
In the following we always consider the sea quark mass in the so-called $p$-regime.
For the other scales we consider $m\sim\mu_v\sim a\sim1/L\sim O(p^2)$.
The result of the calculation~\cite{Necco:2011vx} in terms of the PCAC quark mass $m_{\rm PCAC}$ is given by
\begin{eqnarray}
[\rho_Q(\lambda,m_{\rm PCAC})+\rho_Q(-\lambda,m_{\rm PCAC})]_{NLO} &=& 2\left[\rho_Q(\lambda,m_{\rm PCAC})\right]_{NLO,cont} \\ 
&+& \frac{2\Sigma\lambda}{\pi\sqrt{\lambda^2-m_{\rm PCAC}^2}}\Bigg[\frac{m_{\rm PCAC}^2\Delta}{\lambda^2-m_{\rm PCAC}^2}+\frac{16\hat{a}}{F^2}W_6\Bigg]\,.
\label{eq:rhoQ_nlo}
\end{eqnarray}
In these formula $\Delta=-\frac{16\hat{a}}{F^2} \left(\frac{W_8}{2} +\frac{W_{10}}{4} +\frac{\hat{a}W_8'}{M^2_{ss}}\right)$,
$M_{ss}$ is the pseudoscalar meson mass (made of two sea quarks),
$\Sigma$ and $F$ are the LO LECs, $\hat{a}=2W_0a$ and $W_0$ as the others $W$s 
are the LECs parametrizing O($a$) and O($a^2$) effects. 
Details on the calculation and a discussion on the applicability of this formula
can be found in refs.~\cite{Necco:2011vx,Necco:2011jz}. 


Potentially even with Wilson fermions one can use the spectral density, or equivalently 
the renormalization group invariant (RGI) mode number~\cite{Giusti:2008vb}
\be
\nu\left(\Lambda,m\right) = V \int_{-\Lambda}^\Lambda d~\lambda \rho_Q\left(\lambda,m\right)\,,
\label{eq:mn}
\ee 
to determine the chiral condensate.
To test our formula we compared the resulting mode number~\eqref{eq:mn} 
with the numerical data published in~\cite{Giusti:2008vb}.
We have fixed $F=90$ MeV and the renormalisation scale 
$\mu=m_\pi$; we have performed a global fit at all the $3$ masses available and all the values 
of $\Lambda_R$ with $3$ fit parameters: $\Sigma$, $\Delta$ and 
$\overline{L}_6$.\footnote{In this proceedings with $\Sigma$ we denote its value
renormalised in the $\overline{MS}$ scheme at a scale of $2$ GeV.}
From the global fit we obtain
\be
\Sigma^{1/3} = 266(7) \, {\rm MeV}\,, \qquad \Delta = - 0.62(80)\,, \qquad \overline{L}_6 = 6(1)\,.
\ee 
The numerical data and our global fit are shown in the right plot of fig.~\ref{fig:rhonu_p}.
For $\Sigma$ we obtain a perfectly consistent result with Giusti and L\"uscher~\cite{Giusti:2008vb} without
performing any chiral extrapolation. 
We have also performed a fit fixing $F=80$ MeV obtaining compatible results within errors.
Lattice determinations of the chiral condensate using the mode number can be found in~\cite{Giusti:2008vb,Fukaya:2010na,Cichy:2011an}.
\vspace{-0.3cm}
\section{Cutoff effects close to the threshold}
\vspace{-0.4cm}
\noindent To improve the theoretical description of the spectral density of the
Wilson operator for $\lambda \simeq m$ one needs to consider two important points.
The first one is that finite size effects diverge for $\lambda \rightarrow m$~\cite{Necco:2011vx} 
and the second one is that the when $\lambda \simeq m$ the power counting for the valence quark masses
need to be reconsidered, i.e. the treatment of the cutoff effects in a perturbative fashion might not
be adequate. 

To overcome this difficulties we opt for the following power counting
\be
m\sim O(p^2),\quad m_P=\sqrt{m_v^2+\mu_v^2} \sim O(p^4), \quad 1/L,1/T \sim  O(p)\,, \quad a\sim O(p^3)\,,
\label{eq:power}
\ee
that implies cutoff effects affecting the spectral density at NLO order.  
The framework is the so-called \emph{mixed Chiral Effective Theory}~\cite{Bernardoni:2007hi}, 
where some masses obey the $p$-regime counting, and others are in the epsilon-regime.
As an intermediate step of our calculation we introduce a $\theta$-term solely in the mass-term of the action as follows
\be
\mathcal{L}_2 = \frac{F^2}{4}{\rm Tr}\left[\partial_\mu U\partial_\mu U^\dagger\right] -
\frac{\Sigma}{2}{\rm Tr}\left[U_\theta^\dagger U(x)^\dagger\mathcal{M}+ \mathcal{M}^\dagger U(x) U_\theta\right]-
\frac{\hat{a}F^2}{4}{\rm Tr}\left[U+U^\dagger\right]\,,
\ee
where the mass matrix in the replica formalism~\cite{Damgaard:2000gh} is
\be
\mathcal{M} = \mathcal{M}^{\dagger}={\rm diag} (\underbrace{m,\ldots,m}_{N_s}, 
\underbrace{m_v+i\mu_v\tau^3,\ldots,m_v+i\mu_v\tau^3}_{N_r})\,,\,\,\,\,\,\,\,\,\,\,\,\,\,\,\,\,\,\,
U_\theta={\rm diag}(e^{\frac{i\theta}{N_s}\mathbbm{1}_s}, \mathbbm{1}_r) \,.
\label{eq:massterm}
\ee
While in the continuum it is not important how the $\theta$-term is introduced in the 
parametrization of the $U$-field~\cite{Bernardoni:2007hi,Bernardoni:2008ei},
adding a $\theta$-term in the sea sector only, becomes
relevant at finite lattice spacing. With this choice, even with a $\theta$-term in the action, 
we can reabsorb the leading O($a$) cutoff effects in a redifinition
of the quark mass, because
with our choice of power counting~\eqref{eq:power}, in the sea sector the leading O($a$) effects appear
at NNLO, i.e. the sea quarks are effectively in the continuum up to higher order corrections.
With this particular choice of power counting and parametrization of the $\theta$-term we can achieve, 
as in the continuum~\cite{Bernardoni:2007hi,Bernardoni:2008ei}, 
a factorization of the partition function for the zero and non-zero modes.
The periodicity in $\theta$ of the chiral Lagrangian allows us to write the partition function
in standard fashion
\be
\mathcal{Z}(\theta) = \sum_{\nu=-\infty}^{\nu=+\infty} e^{-i \nu \theta}\mathcal{Z}_\nu\,, \qquad 
\mathcal{Z}_\nu = \frac{1}{2 \pi} \int_0^{2 \pi} d \theta~ e^{i \nu \theta} \mathcal{Z}(\theta)\,.
\ee
By performing an exact integration over the constant field $\theta$ one obtains 
\be
\mathcal{Z}_\nu \propto e^{\frac{-N_s \nu^2}{2 z_s}} \int_{U(N_r)} d \overline{U}_0 
\left( \det \overline{U}_0 \right)^\nu {\rm e}^{\frac{m_v \Sigma V}{2} 
{\rm Tr}\left[ \overline{U}_0^\dagger + \overline{U}_0 \right]}, 
\qquad z_s = m V \Sigma\,,
\label{eq:Znu}
\ee
from which one observes that the distribution of $\nu$ is Gaussian
and it is controlled by the sea quarks which are in the $p$-regime.
The computation of the spectral density is now a computation at fixed $\nu$, i.e. 
$\rho= \sum_{\nu} \rho_\nu \frac{\mathcal{Z}_\nu}{\mathcal{Z}}$.
The sum over $\nu$ can be done because we know the weight factor $\frac{\mathcal{Z}_\nu}{\mathcal{Z}}$
(cfr. eq.~\ref{eq:Znu}). We expand the pseudo Nambu-Goldstone field $U(x)$ around the ground state of the theory~$U_V$
\begin{equation}\label{eq:vac_replica}
U_V={\rm diag} (\underbrace{1}_{N_s}, \underbrace{e^{i\tau^3\omega_0}}_{N_r})\,, \quad \sin\omega_0=\frac{\mu_v}{m_P},\;\;\;\cos\omega_0=\frac{m_v}{m_P}\,,
\end{equation} 
and with this parametrization, the mass term in the chiral Lagrangian becomes 
like in the untwisted case, with a degenerate polar mass $m_P$ in the valence (replicated) sector.

We compute the pseudoscalar valence condensate (cf. eq.~\eqref{eq:disc_P3}) in the chiral effective theory,
\be
\langle P^3_v \rangle =\frac{1}{V}\Big{\langle} \frac{\partial}{\partial J} \int d^4x \mathcal{L}(x) \Big{\rangle} |_{J=0},
\ee
where now the source term has the following form
\be
\mathcal{M}\rightarrow \mathcal{M}_J=\mathcal{M}+J\hat{\tau}^3, \;\;\;\;\mathcal{M}^\dagger\rightarrow  \mathcal{M}^\dagger_J=\mathcal{M}^\dagger-J\hat{\tau}^3\,, \qquad
\hat{\tau}^3={\rm diag} (\underbrace{0}_{N_s}, \underbrace{\underbrace{\tau^3}_{2},0,\ldots,0}_{N_r}).
\ee
The final result of the calculation can be written as
\begin{eqnarray}
\langle P^3_v \rangle &=&\frac{\Sigma_{\rm eff}}{2}F_1
 + 8B\hat{a}W_6F_2+2B\hat{a}W_8F_3\\
&+&\frac{\Sigma}{2}\left[ 4B\hat{a}W_6N_smV
  F_2+\hat{a}^2VW_6'(F_4+4N_sF_2)
+\hat{a}^2VW_7'F_5+\hat{a}^2VW_8'F_6\right],\nonumber
\end{eqnarray}
where $F_{i=1..6}$ are integrals over zero modes and $\Sigma_{\rm eff}$ is an effective chiral condensate
\begin{wrapfigure}{r}{0.5\textwidth}
  \vspace{-30pt}
  \begin{center}
    \includegraphics[width=0.48\textwidth]{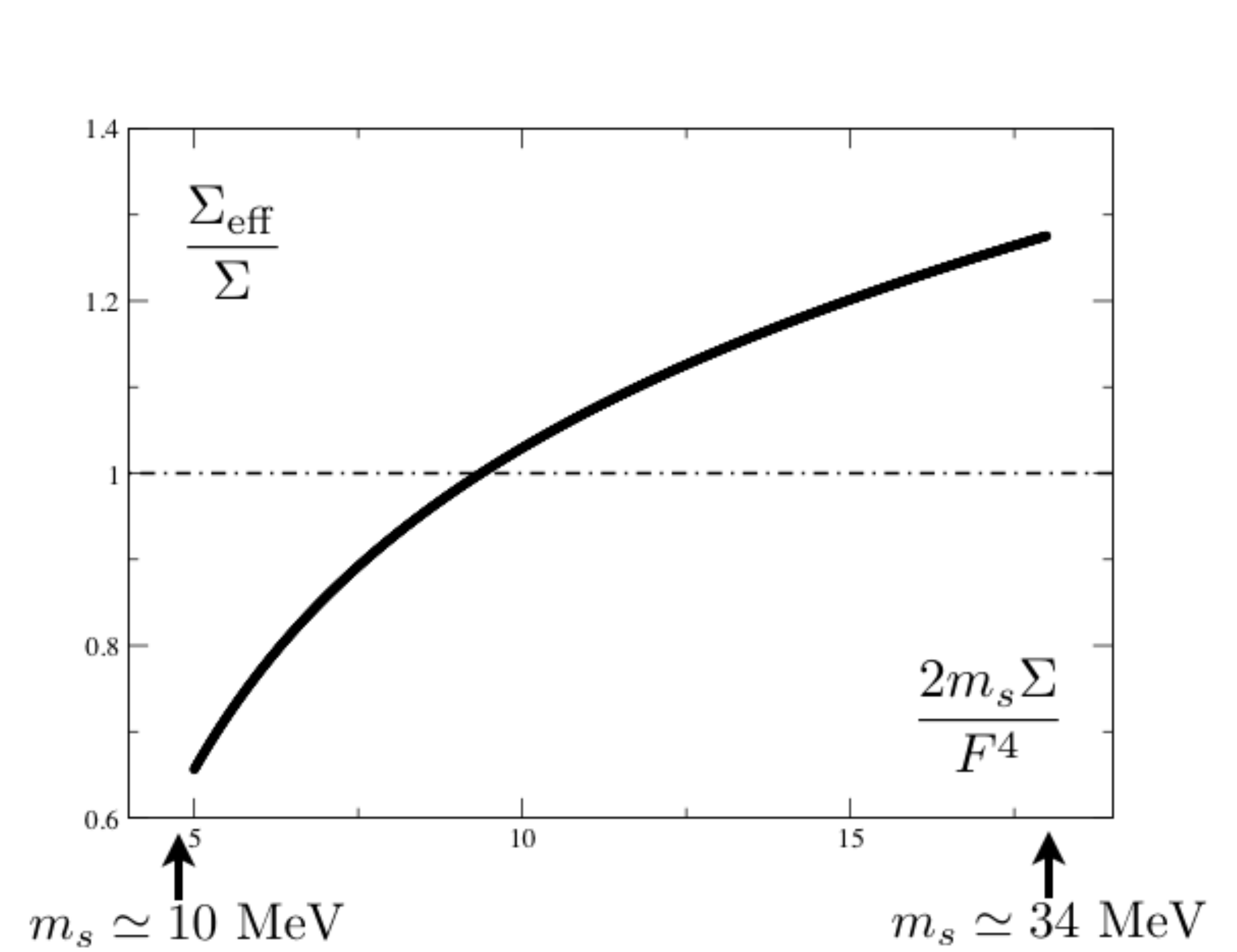}
  \end{center}
  \vspace{-20pt}
  \caption{Significance of the NLO corrections for $\Sigma_{\rm eff}/\Sigma$ in a typical
range of sea quark masses for dynamical lattice QCD computations.}
  \vspace{-10pt}
\label{fig:sigma_eff}
\end{wrapfigure}
which includes NLO corrections from the sea quarks and the lattice spacing
\begin{equation}
\Sigma_{\rm eff}=  \left(\Sigma_{\rm eff}\right)_{\rm cont} + 16\Sigma\frac{\hat{a}W_6}{F^2}\,.
\ee
The explicit formula for $\left(\Sigma_{\rm eff}\right)_{\rm cont}$ 
can be found in~\cite{Bernardoni:2010nf}.
In fig.~(\ref{fig:sigma_eff}) we show the impact of the NLO corrections to the 
chiral condensate~\cite{Bernardoni:2010nf}. One observes that in the typical range of quark masses
for dynamical simulations the relative corrections induced by the presence
of sea quarks can reach $30\%$. This is a warning in case one would like to extract the
chiral condensate from fits of mode numbers where the effect of the sea quarks have been 
neglected.
To compute the zero-modes integrals we switch to the
supersymmetric formulation. All integrals at fixed $\nu$ can be computed by
derivating the graded $SU(2|2)$ partition function~\cite{Fyodorov:2002wq,Splittorff:2002eb} with respect to
appropriately chosen sources. The integrals have been computed and cross-checked. The final step to 
determine the formula for the spectral density is the calculation of the discontinuity along the imaginary axis
in the twisted valence mass plane (cfr.~\eqref{eq:disc_P3}). The analysis of the final result is in progress.
We conclude observing that the 
effects of the sea quarks are twofold. They change the absolute
normalization by introducing an $m$ dependence in $\Sigma_{\rm eff}$ and they control the distribution of $\nu$,
which, we remark, stays Gaussian only because the sea quarks are in the $p$-regime.

\bibliographystyle{h-elsevier}
\bibliography{wbc}

\end{document}